# An Active High Impedance Surface for Low Profile Tunable and Steerable Antennas

Filippo Costa, *Student Member, IEEE*, Agostino Monorchio, *Senior Member, IEEE*, Salvatore Talarico, Fabio Michele Valeri

***Abstract*** **— In this letter, an approach for designing a tunable and steerable antenna is presented. The antenna model is based on a wideband bow-tie radiating element mounted above an active Artificial Magnetic Conductor (AMC). The AMC geometry consists of a Frequency Selective Surface (FSS) printed on a thin grounded dielectric slab in which some chip-set varactor diodes are placed between the metallic elements and the backing plane through vias. The resulting antenna can be tuned over the S-Band by simply changing all varactor capacitances through an appropriate biasing voltage. Moreover, this structure can operate a beam scanning over each working frequency by applying an appropriate biasing voltage to the active elements of the AMC surface in accordance to leaky radiation principles. The low profile active antenna is characterized by an overall thickness of 5.32 mm, which corresponds to approximately λ/24 at the centre of the operating band.**

## I. INTRODUCTION

In the last years, the interest in development of tunable and steerable antennas for communications, electronic surveillance and countermeasures has been increased. It is due to the capability of these antennas in adapting their properties to achieve selectivity in frequency, bandwidth, desired polarization, gain and direction of beam. In particular, preliminary studies have been carried out to demonstrate electronic tunability for different antenna structures.

In literature, the dynamic behaviour is addressed either by using varactor diodes [1], or by employing electrically [2] and magnetically tunable substrates and by using of barium strontium titanate (BST) and ferrite materials [3].

Tuning of printed dipoles or slot antennas has also been considered since they share the same advantages of portability, low profile characteristic and compatibility in integration with other monolithic microwave integrated circuits (MMICs). Kawasaki and Itoh [4] presented a slot tunable antenna loaded with reactive FET components. Several interesting approaches were presented by Maddella *et al.* [5], who created a tunable coplanar rectangular patch antenna by using a MEMS varactors. RF MEMS switches are also used by Huff and Bernhard [6] to reconfigure the radiation patterns of a resonant square spiral microstrip antenna between endfire and broadside over a common impedance bandwidth. Yang and Rahmat Samii [7] showed the possibility of obtain a circular polarization diversity by mounting a switching diode at the center of a slot cut on a patch antenna. Sievenpiper presented electronically steerable leaky wave antenna by covering a metal ground plane with a periodic surface texture, in which varactor diodes are incorporated [8]. In this way, a biasing voltage controls the resonance frequency of a tunable high impedance surface. Another interesting approach is presented by Bray and Werner in [9], where a broadband open-sleeve dipole antenna is mounted above a tunable AMC, resulting in a tunable low-profile antenna. An analytical model of active high impedance surfaces was recentely presented by Luukkonen *et al.* [10].

In this paper we show the procedure for obtaining both the tunability and the beam steering of the antenna by simply using one radiating element placed in near field of a tunable AMC. Moreover, a technological approach for simplifying the active AMC biasing is addressed in the next paragraph. Computed and measured results are compared.

## II. AMC DESIGN

The unit cell of the Artificial Magnetic Conductor is a square metallic patch of 1.6 cm periodicity. The gap between adjacent patches is equal to 0.54 cm. By employing this simple shape, both a simple biasing and a wideband AMC behavior can be reached [8], [11]. Diodes on FSSs can be introduced either between neighbor patches [12] or by connecting active elements between each FSS element and the ground plane. The former choice calls for a complex feed network in which several couple of wires must reach each diode. Moreover, in this case, a single row of the FSS can not be loaded by the active elements. In the proposed configuration, the varactors can be simply fed by applying a positive voltage to the FSS and a negative one to the ground plane. The Abrupt Tuning Varactors MTV-30-05-08 are connected between the backing plane and the patches through metalized vias. The patches are arranged in a 5 by 5 elements square grid connected each other in five lines. The elements of each column are electrically connected and all lines have a thin strip-line with a resistor of 10 KΩ at their ends and a pad for biasing (Fig. 1). The FSS is printed by lithographic technique on a 1.52 mmm Taconic RF60 dielectric substrate backed by a PEC. It has to be pointed out that, due to the package effects and radio-fre-

F. Costa, A. Monorchio, S. Talarico are with the Department of Information Engineering, University of Pisa, Via G. Caruso 16, I-56122 Pisa, Italy, (a.monorchio@iet.unipi.it -www.iet.unipi.it).

M.F. Valeri is with is with Microwave Communications, Pisa, Italy (www.microcomm-it.com).

quency resistance of the varactors, particular care is required for choosing the suitable active elements.

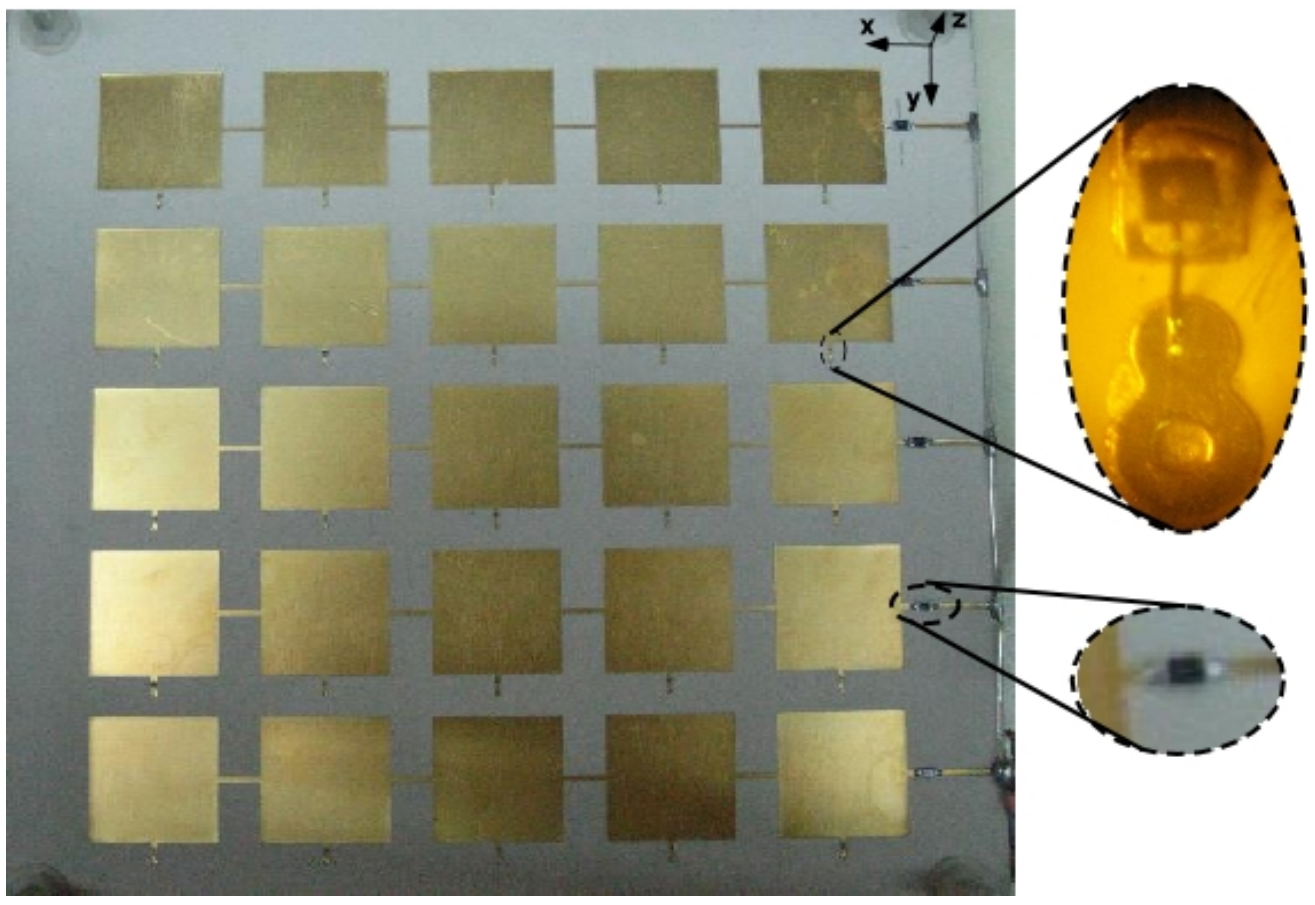

Fig. 1 - Geometry of AMC surface with a detail of a diode and a resistance of the polarizing network.

In particular, shunt capacitance directly across the terminals of the diode and series resistance represent the main problems. The first has been overcome by using a chipset varactor soldered to the FSS by wires bonding [13]. The series resistance takes into account the diode resistance and the electrical contact. It is kept down to 0.3 Ω by using an high Q varactor. The structure has been analyzed by using Ansoft HFSS v.10 , where the active elements are represented by RLC models. This model accounts for the parasitic effects of contacts and package. In Fig. 2, AMC reflection phase is shown as a function of junction capacitance.

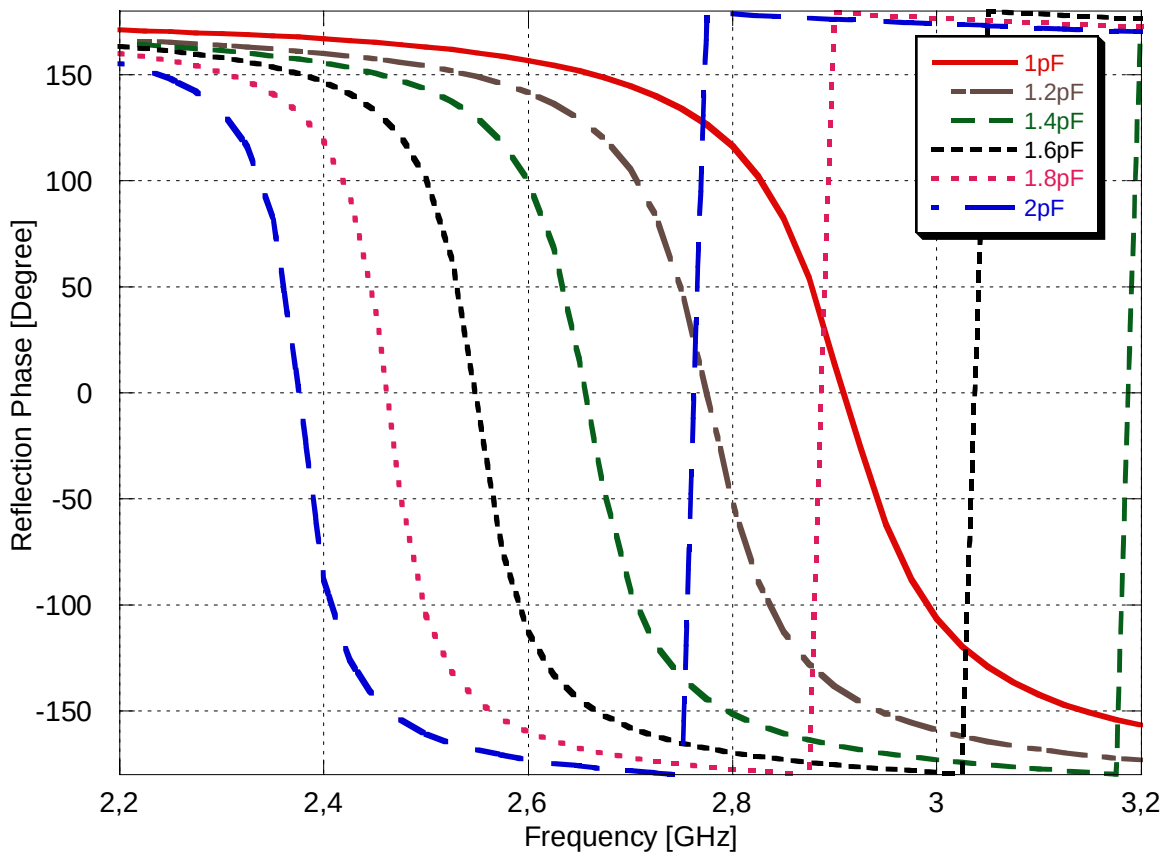

Fig. 2 - Phase response of AMC for different varactor diodes capacitances. The reflection coefficient is calculated for normal incidence with the electric field parallel to the diode orientation.

## III. ANTENNA SYSTEM DESIGN

In order to obtain a narrowband tunable antenna operating in S-Band, a bow-tie radiation element is mounted on a tunable AMC surface at a distance of 3 mm, as shown in Fig. 3. The bow-tie element is fed by two 50 Ω parallel stripes. Physical dimensions of the radiating element are reported in Fig. 3. The antenna is printed on both sides of a 0.8 mm thickness FR4 slab. The bow-tie radiation element is designed to match the working band of the tunable AMC surface. The return loss of the free space bow-tie is under −10 dB in the range (2.68÷3.22) GHz. By changing the working frequency of the AMC, we can control the composite antenna return loss. In particular, when the varactor diode capacitance of the AMC surface changes, the antenna impedance matching frequency is shifted. Antenna return loss as a function of the diode capacitance is reported in Fig. 4.

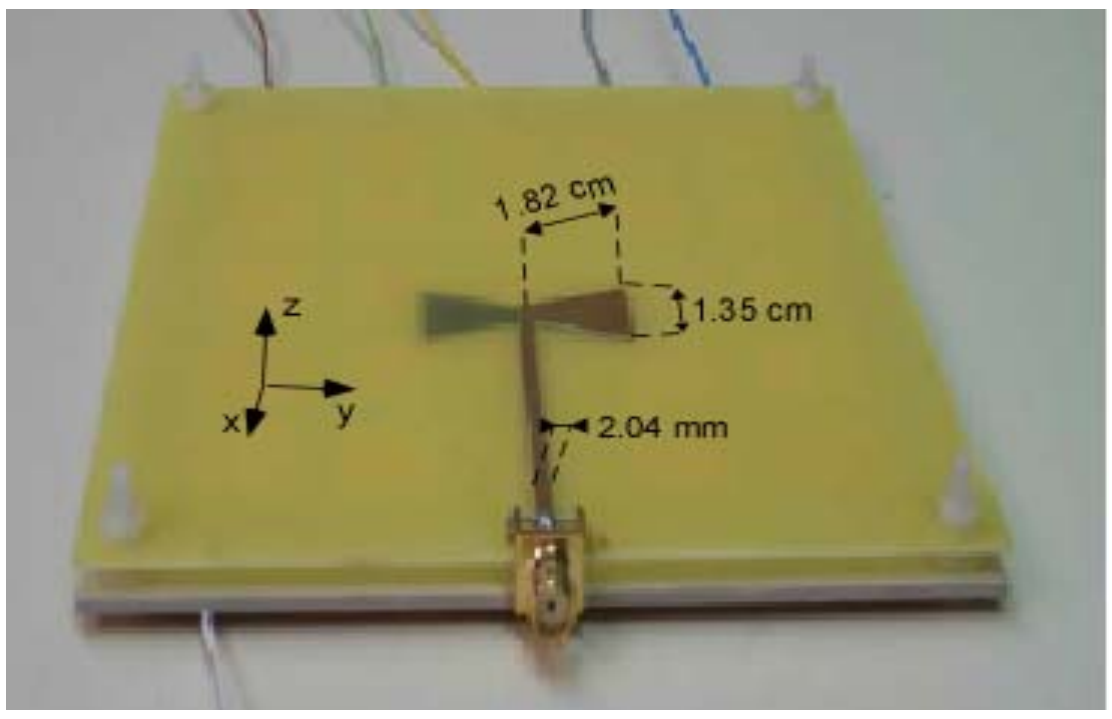

Fig. 3 - A photo of the electronically tunable and steerable antenna.

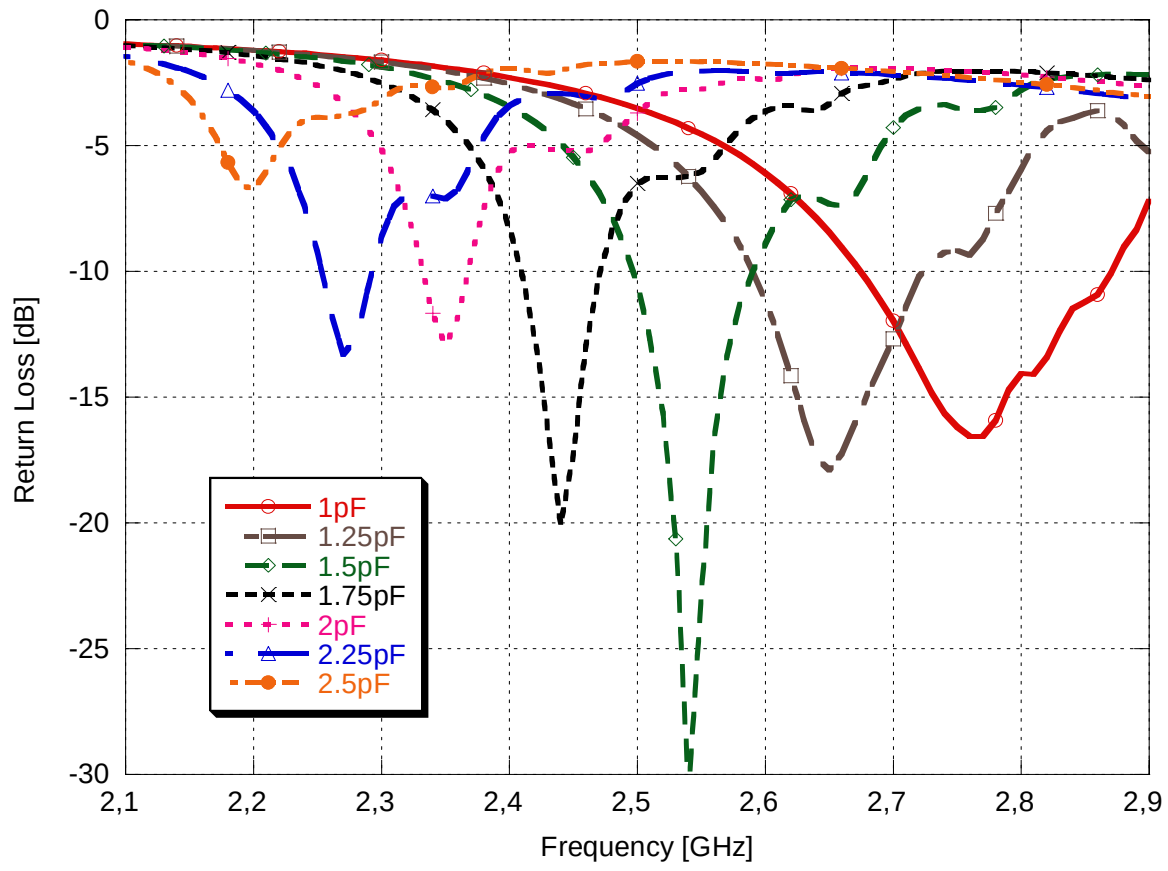

Fig. 4 - Return Loss of the antenna by varying diodes capacitances.

Fig. 5 and Fig. 6 show the antenna radiation patterns on the main planes for different working frequencies. In particular, we focus on the band limits (2.26 GHz – 2.75 GHz) and the centre frequency (2.55 GHz).

One of the main inconveniences of this kind of configuration may result from the occurrence of nonlinearities in active elements which limits the maximum RF power accepted by the antenna in transmitting mode. The close interaction between the bow-tie antenna and the active elements (see Fig. 1) causes an high field value on the central diodes. This radio-frequency power determines a radio frequency voltage that has to be added to the dc polarization voltage. With the increasing of the antenna input power, the polarization voltage oscillates around a medium value. When this oscillation rises above the dc voltage, the diodes become reverse polarized causing a malfunctioning of the active antenna. By evaluating the induced voltage on the diode from the electric field radiated by the bow-tie, we estimate a maximum input power of about 1 W, for the configuration here presented. However, it is worth to point out that this problem can be easily overcome by inserting the diodes behind the ground plane by means of some holes [8].

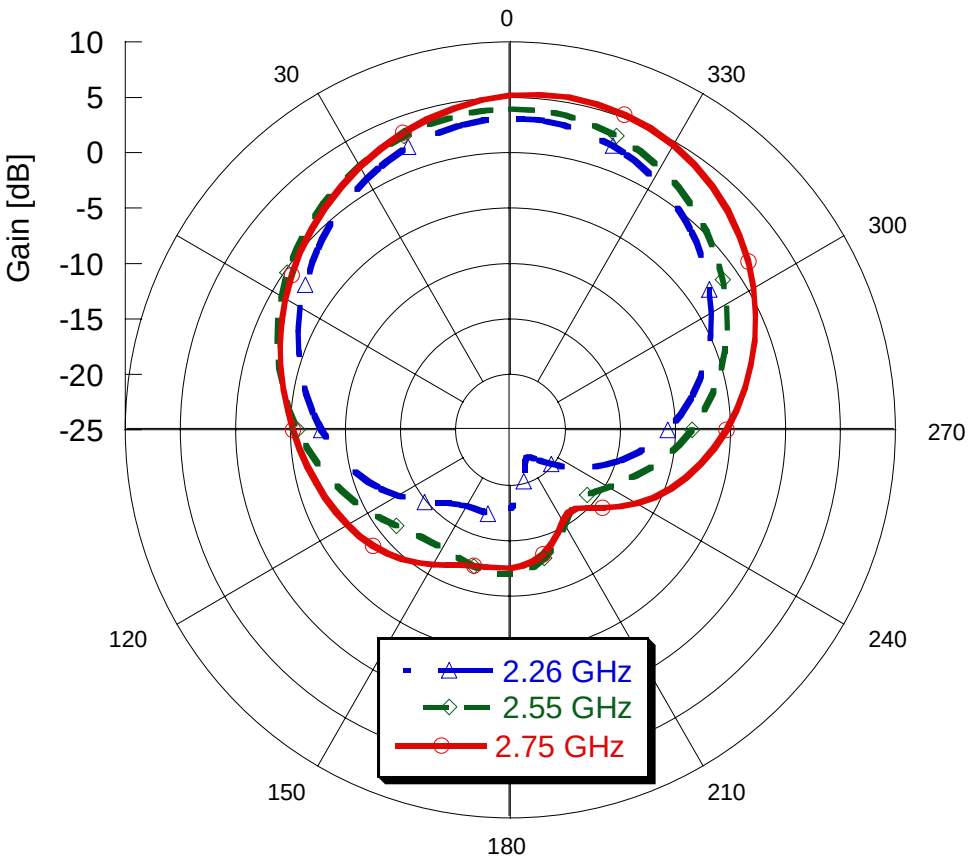

Fig. 5 - Gain for Phi=0° .

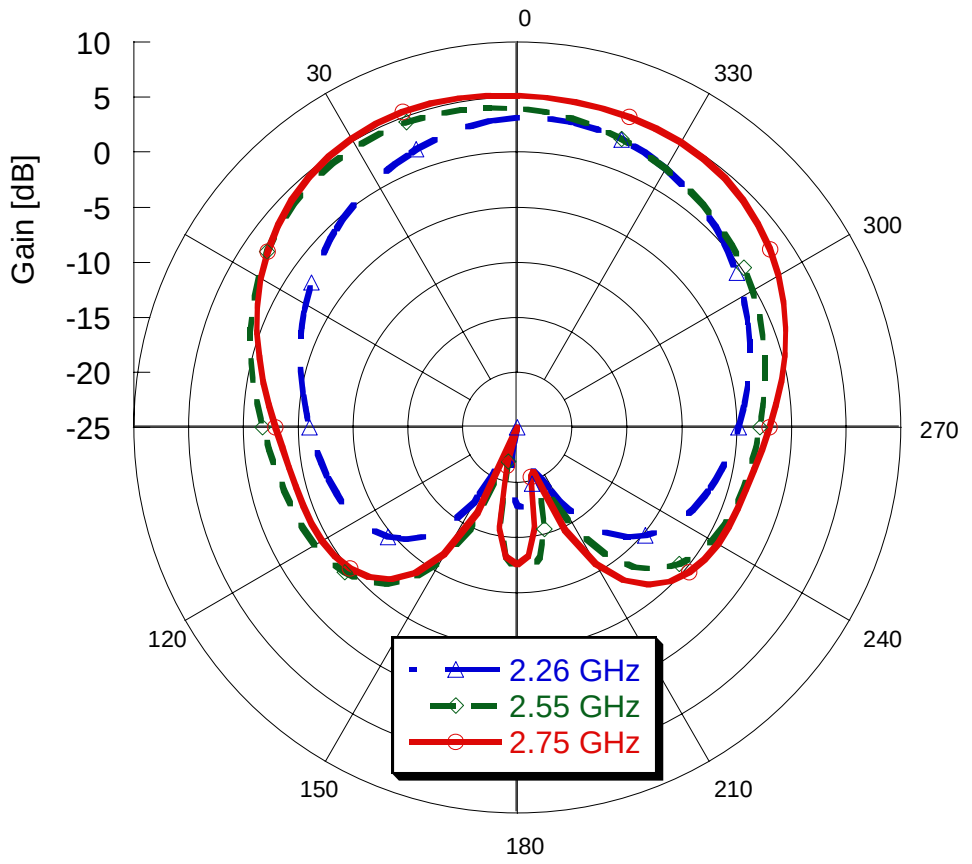

Fig. 6 - Gain for Phi=90°.

## IV. Steering Properties

Radiation properties of such a structure are given by excitation of leaky waves [14]. The presence of these radiating waves can be highlighted by means of the dispersion diagram of the screen. In our case, we are interested to the propagation along the direction where active elements are placed. In particular, $y$-directed leaky wave modes are present in YG zone of dispersion diagram. Radiation or coupling between a space wave and a surface wave requires that the wave vector of the space wave $k_0$ must have a component tangential to the surface that matches the wave vector of the surface wave $k_y$. Radiation cannot occur when $\omega<ck_y$ because there is no angle for which this phase matching condition is satisfied. When $\omega>ck_y$, energy can radiate from the surface into free space towards the angle:

$$\vartheta = \sin^{-1}\left(ck_y/\omega\right) \tag{1}$$

The presence of a forward and a backward wave allows the antenna to perform a beam scanning toward positive and negative angles with respect to the normal direction. The computed dispersion diagram by means of HFSS (Fig. 7) with periodic boundary conditions does not reveal the presence of the backward wave (the presence of a backward wave is not attainable with an infinite structure [15]). Anyway, by computing the finite antenna radiation patterns for different frequencies, this phenomenon is evident.

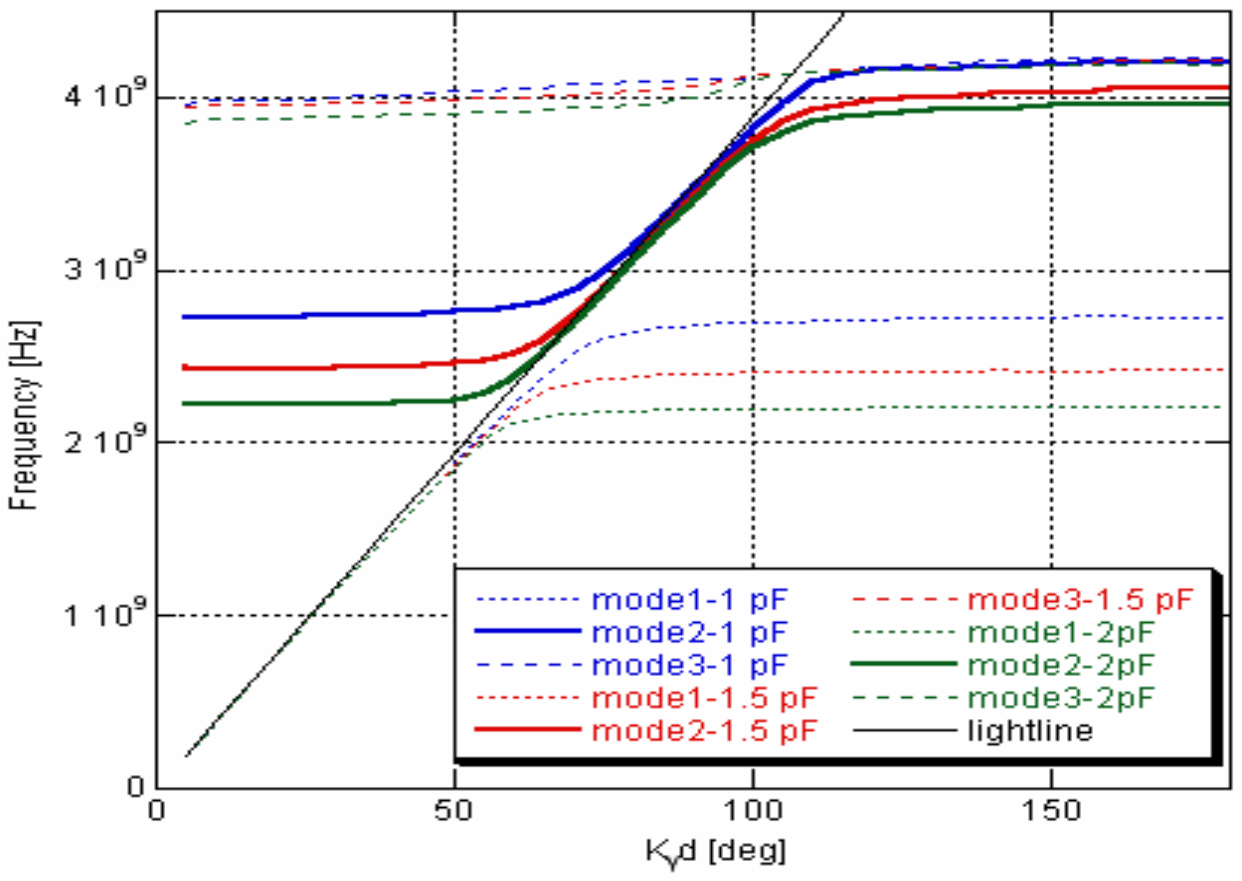

Fig. 7 – Computed dispersion diagram of YG zone for different capacitances.

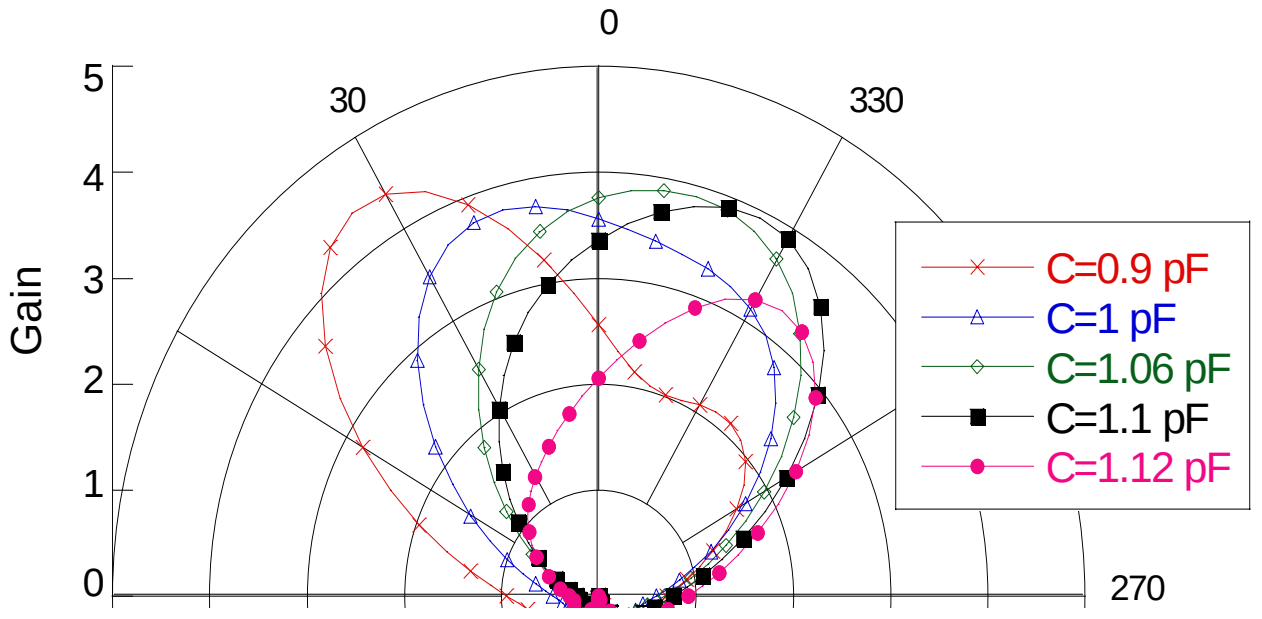

Fig. 8 - Beam steering at 2.76 GHz (gain is in linear scale).

For instance, let us consider the case where all diodes are polarized for a capacitance of 1 pF: starting immediately before the return loss resonance frequency, the radiation pattern is tilted toward positive angles; if we increase frequency, we obtain a broadside pattern in correspondence of the resonance. Moreover, continuing increasing the frequency, the steering direction is now towards negative angles (indeed, this phenomenon is clearly apparent in Fig. 13, in the following section regarding experimental results). By changing diodes polarization, the second mode can be shifted (see Fig. 7) and the same steering can now be obtained in correspondence of a fixed frequency in the same way as described above [8].

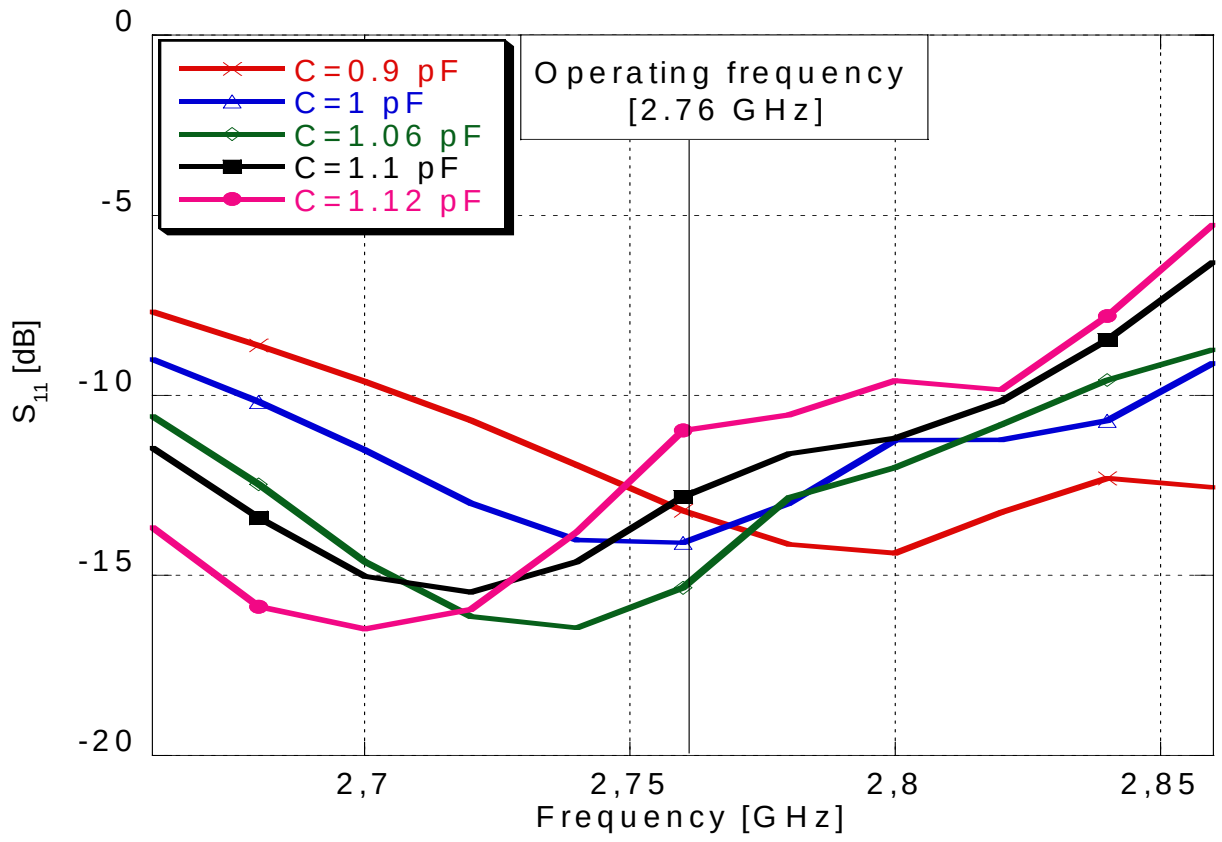

Fig. 9 – Return loss for different capacitance values around 1 pF.

Fig. 8 shows the radiation patterns varying capacitance values around 1 pF. This procedure, due to the rapid variation of the patterns with a little changing of capacitance, allows to preserve the antenna matching as shown in Fig. 9. In this case the working frequency is equal to 2.76 GHz.

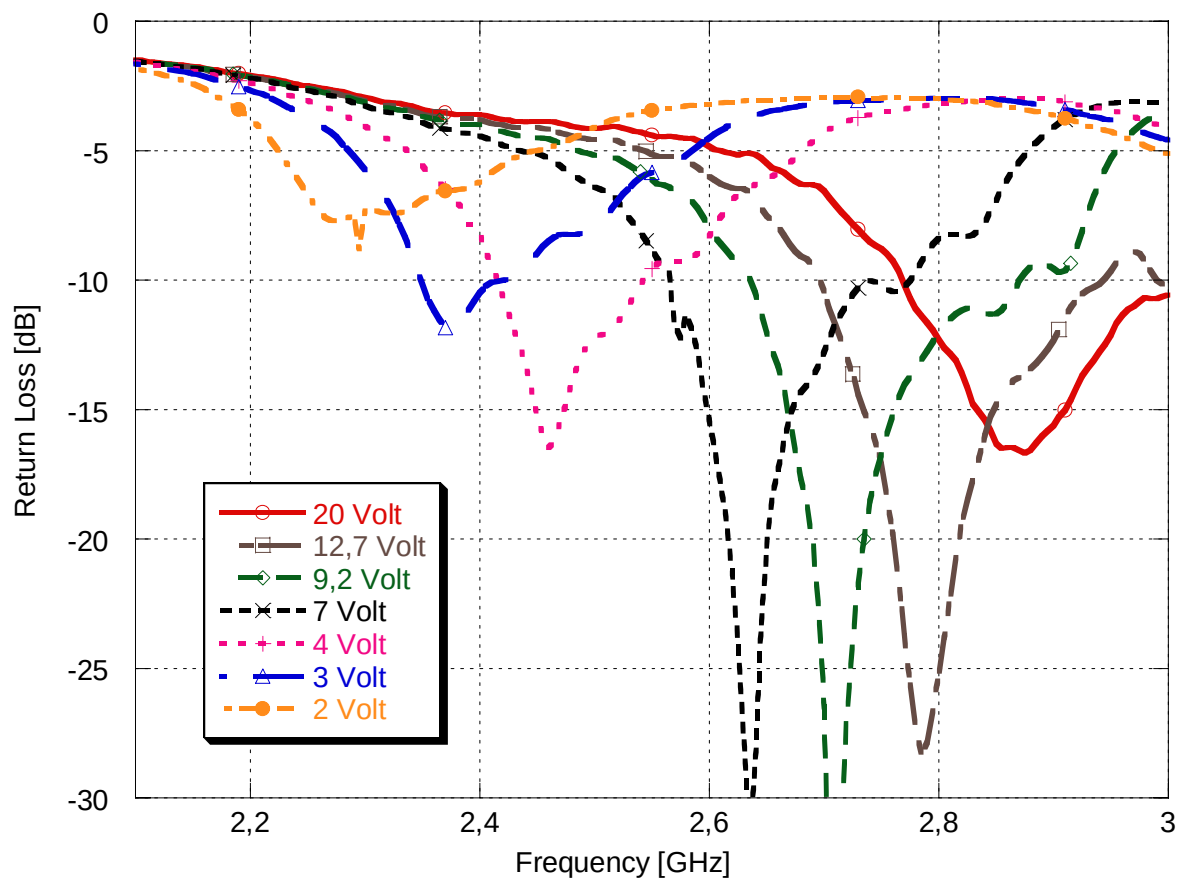

Fig. 10 - Measured resonant frequencies of the reconfigurable antenna by varying diodes dc polarizing voltage.

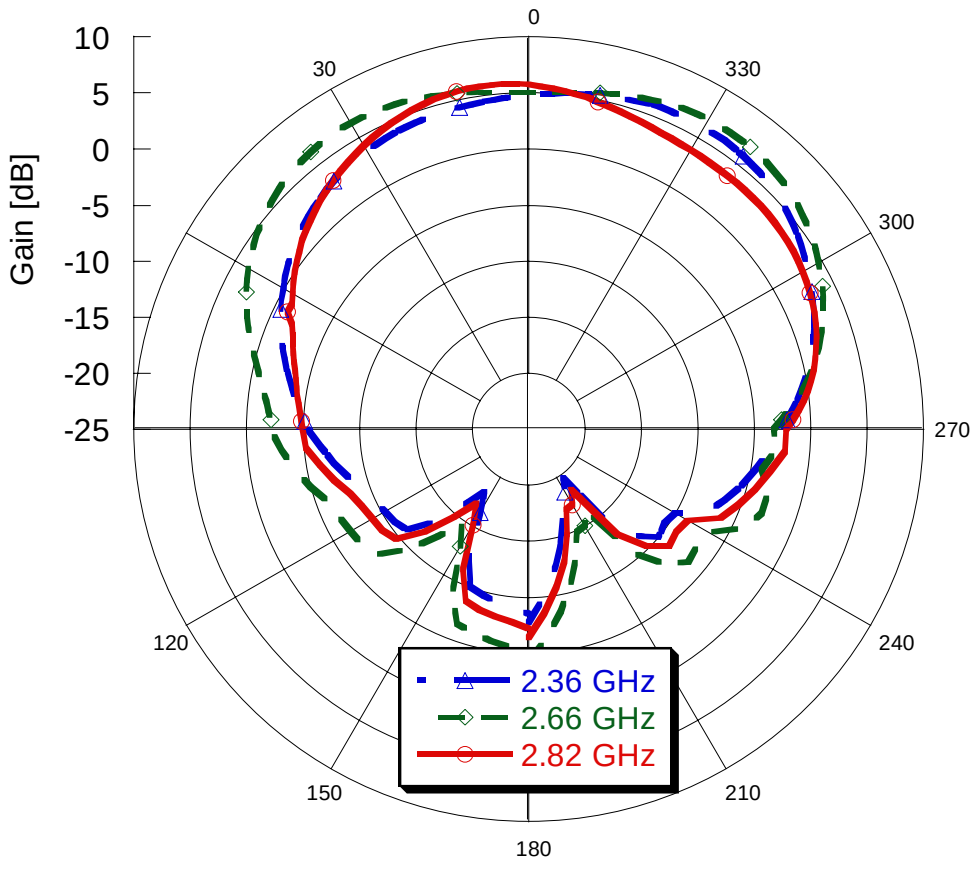

Fig. 11 – Measured radiation Pattern for H-plane (ϕ=0º).

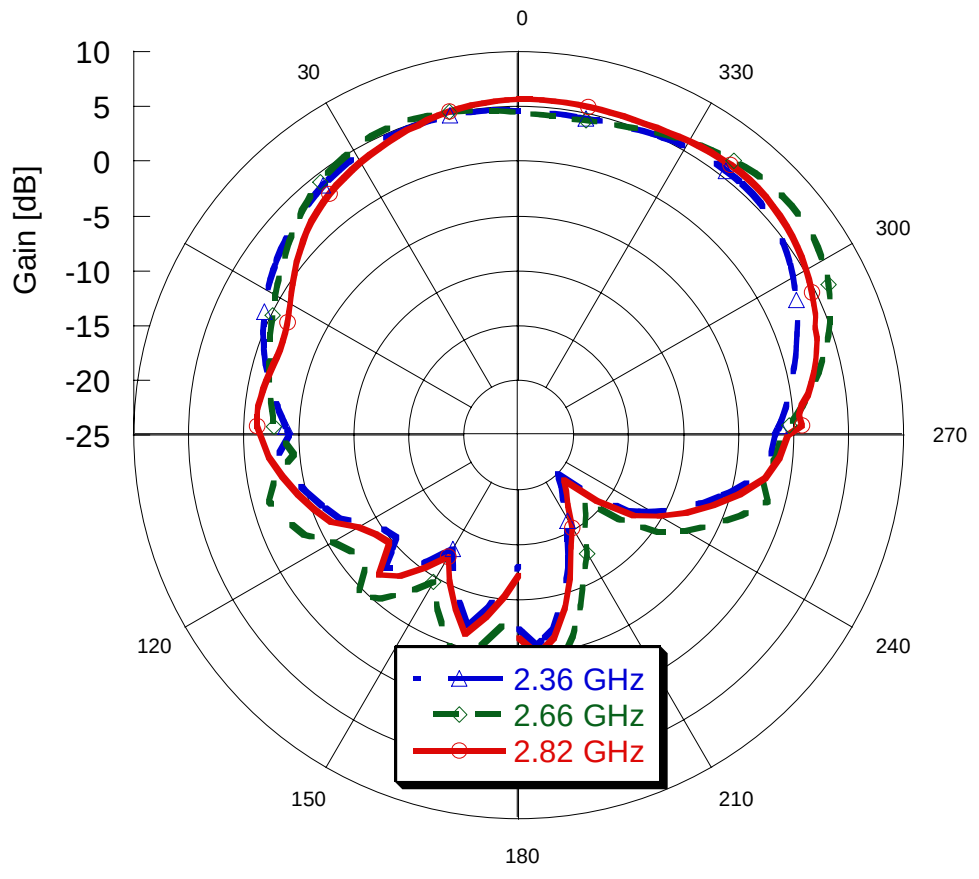

Fig. 12 – Measured radiation Pattern for E-plane (ϕ =90º).

## V. EXPERIMENTAL RESULTS

The antenna return loss was measured for different dc voltages. The biasing voltage for the varactor diodes was provided by a dc voltage source. Voltage values are obtained through the typical relationship between reverse voltage and capacitance of the diode. Measured data are reported in Fig. 10. An excellent agreement between numerical and measured results, but for a small frequency shift, can be observed. In Fig. 11 and Fig. 12 antenna patterns at 2.36 GHz, 2.66 GHz and 2.82 GHz are shown. They correspond to lower band limit, centre frequency and upper band limit, respectively. The steering properties are demonstrated by the measured radiation patterns shown in Fig. 13, obtained at different frequencies and for a fixed dc polarizing voltage equal to 7 V.

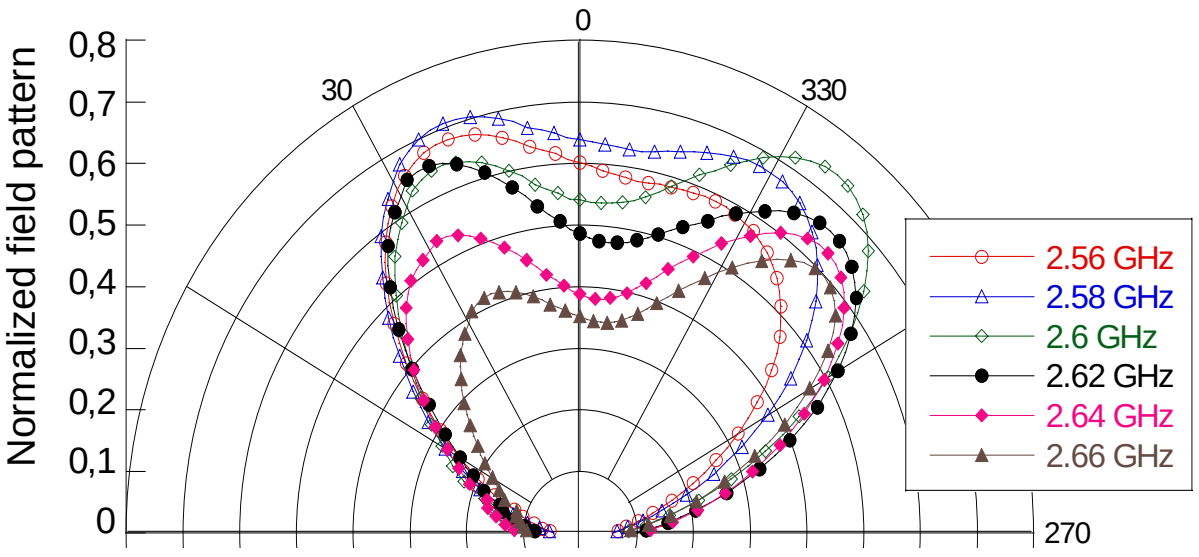

Fig. 13 – Measured normalized radiation patterns on E-plane around resonance frequency with a dc polarizing voltage equal to 7 V (resonance frequency 2.62 GHz).